\documentclass[fleqn,usenatbib]{mnras}

\usepackage{newtxtext,newtxmath}

\usepackage[T1]{fontenc}

\DeclareRobustCommand{\VAN}[3]{#2}
\let\VANthebibliography\thebibliography
\def\thebibliography{\DeclareRobustCommand{\VAN}[3]{##3}\VANthebibliography}

\newcommand{\Msun}{{$M_{\odot}$}}
\newcommand{\Lsun}{{$L_{\odot}$}}

\newcommand{\WH}{$W_{RP} -  W_{K_s}$}

\newcommand{\JKs}{$J-K_s$}

\newcommand{\HI}{H\,{\small I}}

\usepackage{gensymb}
\usepackage{soul}
\usepackage{makecell}
\usepackage{ulem}

\usepackage{graphicx}	
\usepackage{amsmath}	

\usepackage{amssymb}	

\title[Milky Way's warped disc traced by AGB stars]{Milky Way's warped disc traced by AGB stars}

\author[Kushwahaa T. et al.]{Tanya Kushwahaa,$^{1}$\thanks{E-mail: KushwahaaT@cardiff.ac.uk}
Mikako Matsuura,$^{1}$
Jason A. S. Hunt,$^{2}$
Daisuke Kawata,$^{3}$
Roger Wesson,$^{1,4, 5}$  
and  \newauthor
Timothy A. Davis$^{1}$
\\
$^{1}$Cardiff Hub for Astrophysics Research and Technology (CHART),
School of Physics and Astronomy, Cardiff University, 
The Parade, Cardiff CF24 3AA, UK\\
$^{2}$
School of Mathematics \& Physics, University of Surrey, Stag Hill, Guildford, GU2 7XH, UK \\
$^3$
Mullard Space Science Laboratory, University College London, Holmbury St. Mary, Dorking, Surrey, UK \\
$^4$
Department of Physics and Astronomy, University College London, Gower Street, London WC1E 6BT, UK \\
$^{5}$ 
Department of Physics, Maynooth University, Maynooth, Co. Kildare, Ireland \\
}
\date{Accepted XXX. Received YYY; in original form ZZZ}
\pubyear{2025}

\begin{document}
\label{firstpage}
\pagerange{\pageref{firstpage}--\pageref{lastpage}}
\maketitle

\begin{abstract}
While the presence of the Galactic warp has long been established from observations of \HI\, gas, the \textit{Gaia} measurements of over 1 billion stars with parallaxes have enabled much more detailed studies using stellar populations. Here, we demonstrate that asymptotic giant branch (AGB) stars, an evolved phase of low- and intermediate-mass stars, can serve as an effective tracer of the Galactic warp.
We use two distinct AGB populations: C-rich AGB stars, representing stars of about 1~Gyr in age with main-sequence masses of 2--2.5~\Msun, and intermediate-mass (3--5~\Msun) O-rich AGB stars, corresponding to ages of 100--300~Myr. The downward warp traced by O-rich AGB stars is consistent with that found from Cepheids, which is expected given their similar ages.
The more numerous C-rich AGB stars clearly reveal the Galactic warp over a wide range of azimuthal angles. Their warp appears to reach larger amplitudes than that of Cepheids across azimuthal angles. Our results show that C-rich AGB stars, together with intermediate-mass O-rich AGB stars, provide new constraints on the Galactic warp at intermediate stellar ages, offering a new insight into the stellar age and warp amplitude relation.

\end{abstract}

\begin{keywords}
Galaxy: structure -- Galaxy: disc -- stars: AGB and post-AGB
\end{keywords}



\section{Introduction}

The large-scale warp of the outer Galaxy has been recognised for more than seven decades. Commonly used tracers of the warp include \HI\, gas \citep{Levine2006, Kalberla2007}, Cepheids \citep{Chen2019, Skowron2019sci}, red clump stars \citep{Wang.2020xch}, and star clusters \citep{2020A&A...640A...1C}. Here, we show that asymptotic giant branch (AGB) stars provide a powerful tool for revealing the Galactic warp.

AGB stars represent the late evolutionary phase of stars with initial masses of about 1--8 \Msun. They are highly luminous \citep[$10^3$--$10^4$ \Lsun;][]{Vassiliadis:1993p3075, Bloecker.1995} and can therefore illuminate stellar structures in relatively distant systems. Indeed, they have been used to trace the stellar population of the Galactic bar and bulge \citep[e.g.,][]{2023MNRAS.521.2745S, Zhang2024, 2024MNRAS.530.2972S} and the bar in the Magellanic Clouds \citep[e.g.,][]{2019MNRAS.490.1076E}.

Characteristics of AGB stars are their distinct optical spectra: TiO and VO bands are seen in \textit{Gaia} spectra of oxygen-rich AGB stars, while CN bands are present in carbon-rich AGB stars. They also exhibit very red colours in the 2MASS \JKs\ bands \citep[\JKs$>0.8$][]{Frogel1990, Nikolaev2000}.
Consequently, AGB stars can be identified relatively easily in Galactic surveys.

Here, we report that AGB stars can serve as a new tracer of Galactic warps, based on Gaia and 2MASS data. With typical ages of about 1\,Gyr for C-rich AGB stars and 100--300 Myr for intermediate-mass O-rich AGB stars, these populations fill the age gap between previously used stellar warp tracers: Cepheids \citep[$<$400\,Myr; e.g.][]{Skowron2019sci} and tip of the RGB stars \citep[$\sim$2.5 Gyr; e.g. ][]{2019A&A...627A.150R}.

\section{Methods}

We use a combination of \textit{Gaia} XP spectro-photometry \citep{GaiaDR32023} and 2MASS near-infrared photometry \citep{Skrutskie2006}. Two steps of machine learning, combining similarity analysis \citep{Kim2019} and {\sc XGBoost}  \citep{Bethapudi2018} were used to extract O- and C-rich AGB stars from \textit{Gaia} DR3 \citep{Valenari2023, GaiaDR32023} and 2MASS \citep{Skrutskie2006}.  
We adopt geometric distances from \citet{Bailer-Jones2021}, inferred from \textit{Gaia} parallaxes and their associated uncertainties. Only stars with fractional distance uncertainties smaller than 20\% (i.e., $\sigma_d/d < 0.2$) were included  in the analysis. Additionally, we applied a renormalised unit weight error (RUWE) cut of RUWE$\leq$1.4 to exclude stars with potentially unreliable distances, such as unresolved binaries. These selections removes approximately 3\% of the samples from C-rich AGB and intermediate mass O-rich AGB candidates.

\begin{figure}
      \includegraphics[trim={0cm 0cm 0cm 0cm},clip, width=0.47\textwidth]{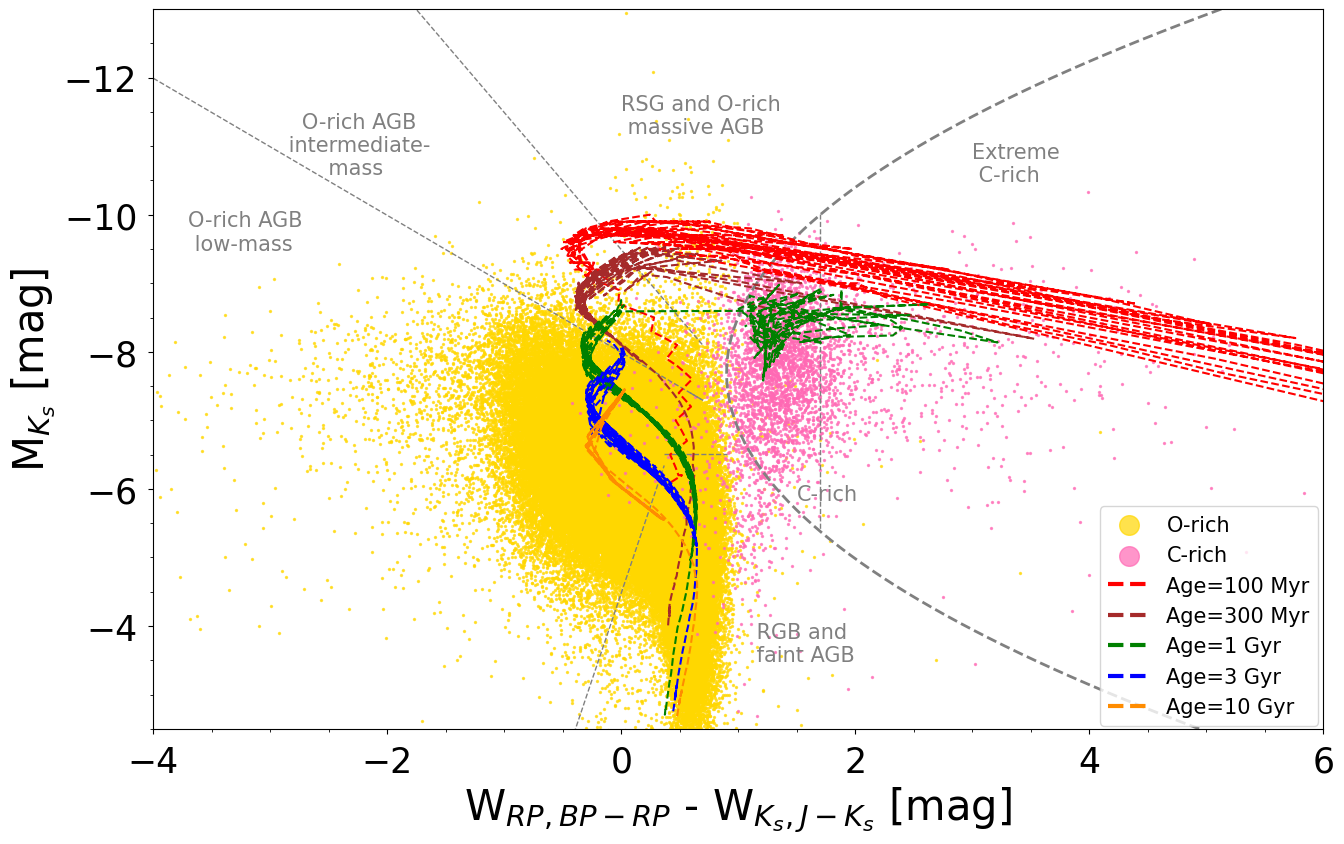}
\caption{\textit{Gaia}-2MASS modified colour-magnitude diagram of the Galactic AGB stars, with the theoretical isochrone from PARSEC-COLIBRI \citep{Bressan2012}. 
Yellow dots show O-rich AGB stars with some red supergiants (RSG) and red giants (RGB) stars, while pink dots show C-rich AGB stars from Kushwahaa et al. (submitted).
A different range of masses, hence, ages are separated well in this diagram.
}
  \label{cm-diagram} 
\end{figure}

The selected AGB candidates are plotted on a modified colour-magnitude diagram (CMD)(Fig.~\ref{cm-diagram}). This CMD \citet{Lebzelter2018} was initially developed to classify AGB stars in the Large Magellanic Cloud, where the distances to the stars are assumed to be the same. With distance estimates from \citet{Bailer-Jones2021}, this CMD can be applied to Galactic AGB stars.
On the x-axis, the Wesenheit index 
$W$ is applied to the colour and is defined as
\WH=
$W_{RP, Bp-RP} -  W_{K_s,J-K_s}$, where
$W_{RP} = G_{RP} - 1.3 \times (G_{BP} - G_{RP})$, and $W_{K_s,J-K_s} = K_s - 0.686 \times (J - K_s)$ \citep{Soszynski2005}. Here, $G_{BP}$ and $G_{RP}$ are \textit{Gaia}'s blue and red photometric magnitudes, and $J$ and $K_s$ are the 2MASS near-infrared magnitudes. 
$W_{RP, Bp-RP} -  W_{K_s,J-K_s}$ is dominated by $W_{RP} -  W_{K_s}$, although it is not a color index in principle.
We describe it as \WH\, in short. 
The Wesenheit function corrects for interstellar extinction under the assumption $R_V = 3.1$, and the coefficients 1.3 and 0.686 are extinction correction factors for $R_V = 3.1$ at these bands.

Fig. \ref{cm-diagram} shows the modified CMD of the Gaia AGB candidates. The theoretical isochrones are also plotted on the CMD diagram. The mass range of these AGB stars is based on Monte Carlo analyses by \citet{Lebzelter2018} and isochrones from PARSEC-COLIBRI \citep{Bressan2012, Marigo2013, Tang2014, Chen2015, Marigo2017, Pastorelli2019, Pastorelli2020}; slightly modified from the MC version to account for Galactic metallicities, which shift stellar colours by approximately 0.2 mag (Kushwahaa et al. submitted). Interstellar medium (ISM) extinction is corrected using the ISM extinction map \citep{Vergely2022}, prior to calculating $K_s$, as well as the colour. We note that for stars beyond $\sim$5~kpc the extinction may be underestimated due to the current limitations of large-scale 3D extinction maps.

The isochrones provide estimates of the main-sequence progenitor masses and current ages of the AGB stars. These derived ranges are summarised in Table\,\ref{tab:agb_stats}. Further details of the analysis are presented in Kushwahaa et al. (submitted).
 
Stars are intrinsically O-rich, and the majority remain so (yellow dots to the left of the long-dashed curve in Fig.\,\ref{cm-diagram}). Stars with initial main sequence masses of around 2--2.5\Msun\ may become C-rich after undergoing several dredge-up episodes \citep{Vassiliadis1993, Karakas2007}. These C-rich AGB stars are shown as magenta dots in Fig.\ref{cm-diagram}.

The isochrone includes emission from circumstellar dust in the very late AGB phase \citep{Groenewegen2006}. Owing to this dust emission, the stars appear redder (\WH$ > 0$ for O-rich AGB stars and \WH$ > 1.8$ for C-rich stars). Since these mass-losing AGB stars, commonly referred to as the tip of the AGB, are short-lived (about 10\% of the AGB phase), their blending into the C-rich regime is minimal (approximately 100 stars; Kushwahaa et al., submitted) and does not significantly affect our Galactic distribution analysis.

\begin{table}
    \centering
    \begin{tabular}{l  r  c  r}
    \hline
    \thead{Classified \\ AGB Region} &  \thead{Number\\ of Stars} & \thead{Mass Range\\ (\Msun}) & Age\\
    \hline
    \textbf{O-rich} & & & \\
    \hline
    RGB and faint AGB     & 160,577 & $\sim$1 & 10 Gyr\\
    Low-mass AGB          & 141,839 & 1--1.5  & 3--10 Gyr \\
    Intermediate-mass AGB$^*$ &   763 & 3--5    & 100--300 Myr \\
    RSG and massive AGB   &     275 & $\sim$9 & 30 Myr \\
    \hline
    \textbf{C-rich} &  & & \\
    \hline
    C-rich (including\\ extreme C-rich) & 3,148 & 2--2.5 & 1 Gyr \\
    \hline
    \end{tabular}
    \caption{Summary of the classified AGB stars plotted in the CMD (Fig.~\ref{cm-diagram}). 
    Their approximate range of masses and ages are estimated from the PARSEC-COLIBRI isochrones in the diagram. The table is taken from Kushwahaa et al. (submitted).}
    \label{tab:agb_stats}
\end{table}

\section{Results}

\begin{figure*}
   \includegraphics[trim={5.cm 1.cm 4.2cm 1.cm},clip, width=0.99\textwidth]{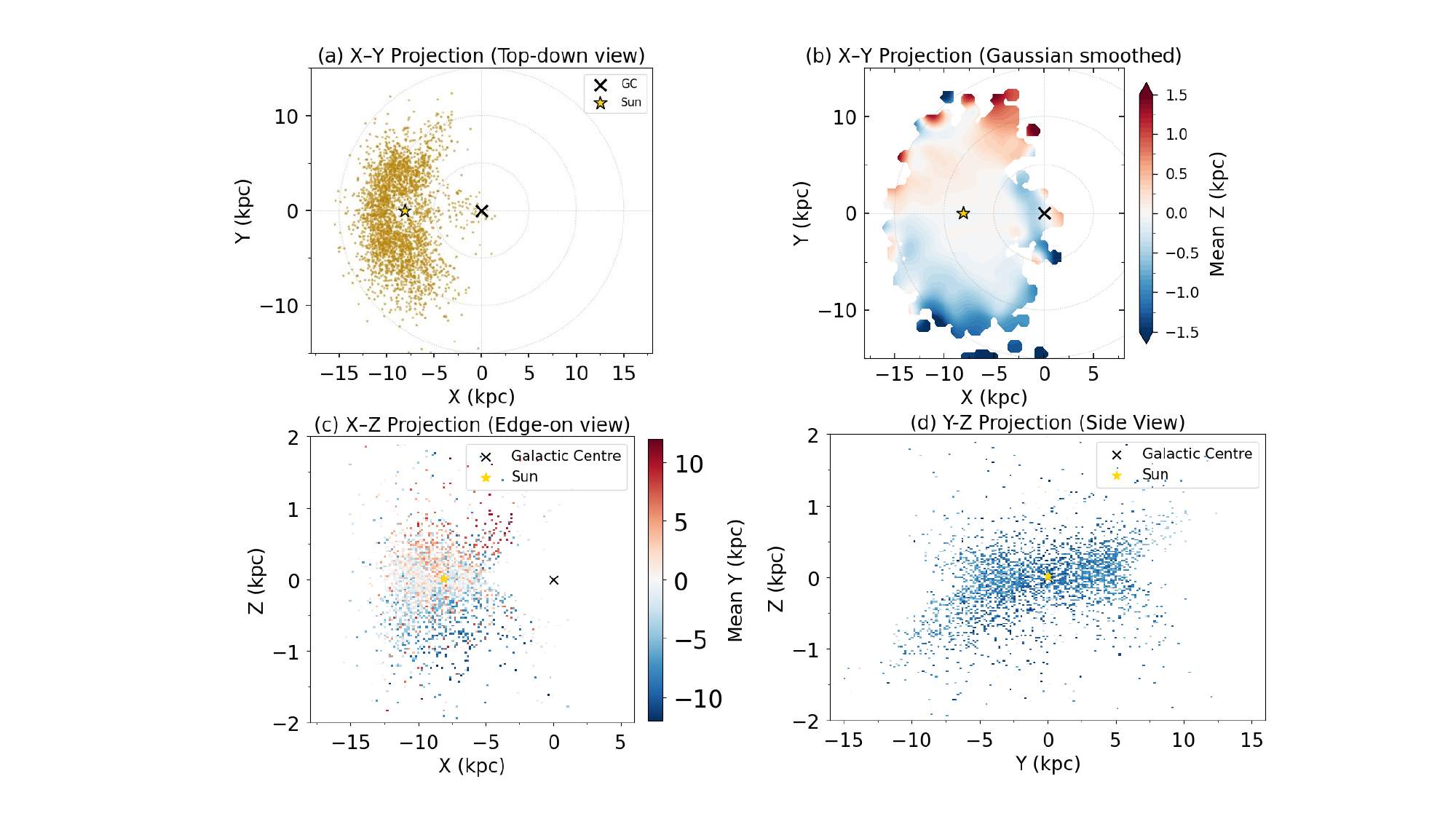}     
\caption{
Galactic distribution of C-rich AGB stars shown in three different views: panel (a) presents a top-down view of the Milky Way, where the individual stars are shown on the left. Panel (b) represents the mean vertical displacement ($Z$) relative to the Galactic plane smoothed using a Gaussian kernel of bandwidth 0.8~kpc to provide a smooth representation of the warp. The Sun is marked with a yellow star symbol. Panel (c) displays a side view, and panel (d) shows the perspective from the Galactic centre toward the Solar System.
Most C-rich AGB stars are located in the outer Galaxy. Panel (b) clearly shows a warp structure with upward displacement (red, positive $Z$) in the positive $Y$ direction and downward displacement (blue, negative $Z$) in the negative $Y$ direction. Panel (d) shows that the warp begins at approximately 6~kpc from the Galactic centre–Solar System line of sight, reaching about $Z \sim 0.7$~kpc and $Z \sim -1$~kpc at $|Y|=14$~kpc.
}
  \label{crich_distributions} 
\end{figure*}

\begin{figure*}
  \includegraphics[trim={5.cm 1.cm 4.2cm 1.cm},clip, width=0.98\textwidth]{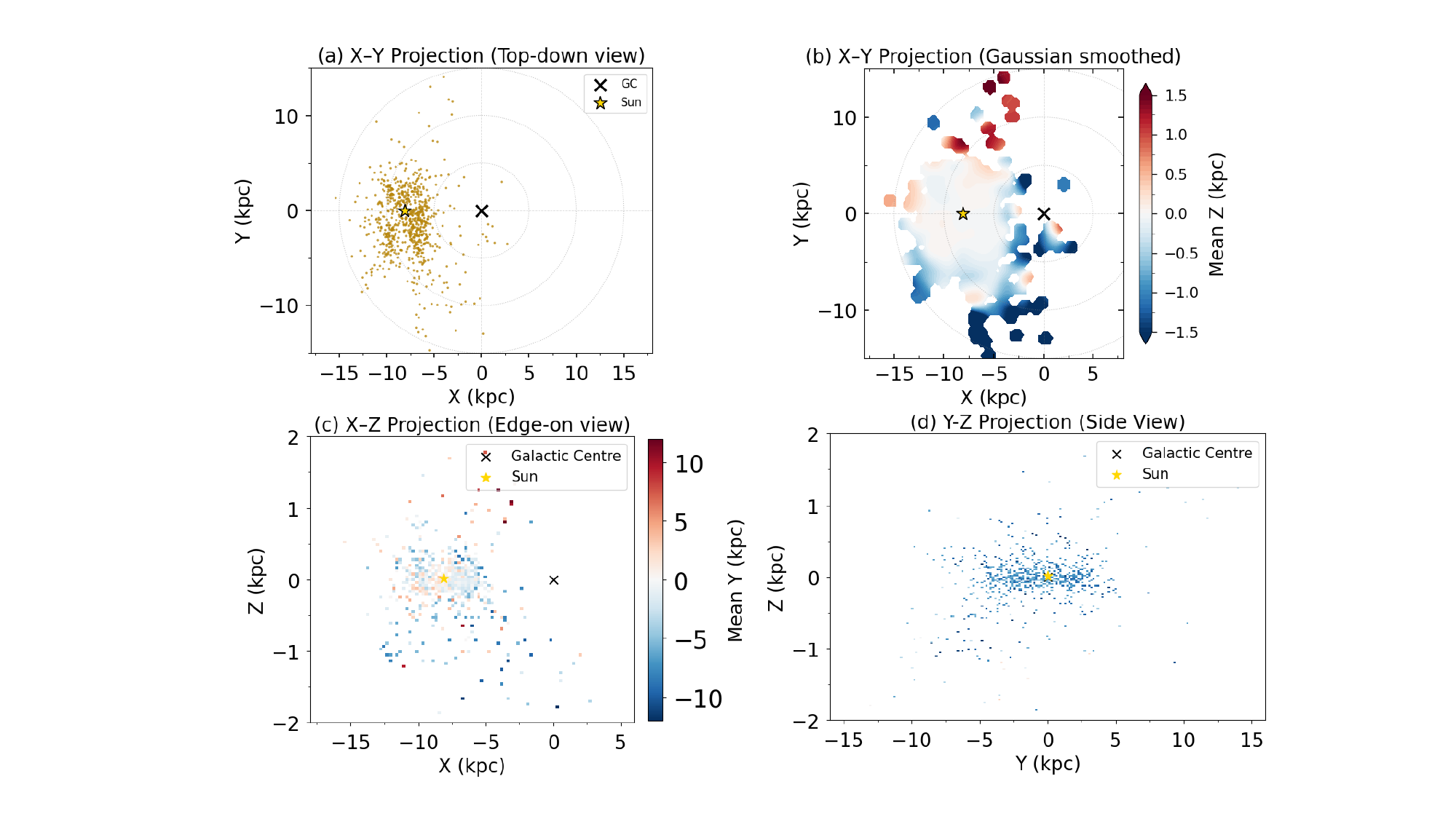}
\caption{Similar to Fig.~\ref{crich_distributions}, but for intermediate-mass O-rich AGB stars. Panel (b) shows a warp pattern similar to C-rich stars, with downward displacement (blue, negative $Z$) at negative $Y$ and upward displacement (red, positive $Z$) at positive $Y$. The side-view in panel (d) displays the warp more clearly, with downward displacement at negative $Y$, while upward displacement at positive $Y$ is weakly constrained due to sparse coverage in the outer disc.}
\label{orich_distributions}
\end{figure*}

We plot the Galactic distributions of two mass ranges of AGB stars, and Fig~\ref{crich_distributions} demonstrates the case of C-rich AGB stars and Fig~\ref{orich_distributions} is for intermediate mass O-rich AGB stars.
The Galactocentric coordinates of the stars are computed, assuming the position of the Sun at (X, Y, Z)=(-8.122, 0, 0.02) kpc \citep{Gravity2018}.

Fig.~\ref{crich_distributions} (a) illustrates the top-down (planar distribution) of individual C-rich AGB stars, showing the underlying spatial coverage of the sample. The majority of these stars are located in the outer Galaxy, beyond the Solar circle, consistent with previous studies of Galactic distributions of C-rich AGB stars \citep[e.g.][]{Jura1989, Lebertre2003}.

Fig.~\ref{crich_distributions}(b) shows the mean vertical displacement (Z) relative to the Galactic plane in a top-down projection. A two-dimensional Gaussian kernel smoothing with a bandwidth of 0.8~kpc is applied on a spatial grid in the X- and Y-directions, so that the transition in Z-displacement is clearly seen.

The colour-coded map clearly shows that stars at $Y<0$ and $R>10$ kpc mainly show a negative displacement from the $Z=0$ plane, while stars at $Y>0$ and $R>10$ display a positive displacement, showing that the C-rich AGB stars clearly reveal the large-scale Galactic warp.

Fig.~\ref{crich_distributions} (c) presents the Galactic distribution of C-rich AGB stars in an edge-on (X–Z) projection, with the colour scale representing the mean Y coordinate at each spatial grid point.
The figure shows a clear contrast, with positive mean $Y$ values at $Z>0$~kpc and negative mean $Y$ values at $Z<0$~kpc, providing an alternative view of the Galactic warp.

Fig.~\ref{crich_distributions} (d) shows that the C-rich AGB stars reveal the large-scale Galactic warp, and provide a clearer view of its amplitude in the Y-Z slice.
The warp begins at a Galactocentric radius of approximately $\pm$6\,kpc and extending out to about 14\,kpc. 

Intermediate-mass (3–5~\Msun) O-rich AGB stars show slightly different Galactic spatial distributions compared to C-rich AGB stars.
Unlike C-rich AGB stars, intermediate-mass O-rich AGB stars are found predominantly in the inner Galaxy (Fig.~\ref{orich_distributions}(a)). This distribution reflects both the metallicity gradient and the fact that younger stellar populations (in this case, O-rich AGB stars with ages of 100--300 Myr) are more concentrated toward the Galactic centre \citep[e.g.][]{Lian.2023}.

The Galactic warp is tentatively detected in intermediate-mass O-rich AGB stars. At $Y<-6$ kpc, there are approximately four times more stars at $Z<0$ kpc than at $Z>0$ kpc (The panel (d) of Fig.~\ref{orich_distributions}).

\begin{figure*}
   \includegraphics[trim={0.5cm 1.cm .9cm 0.8cm},clip, width=0.95\textwidth]{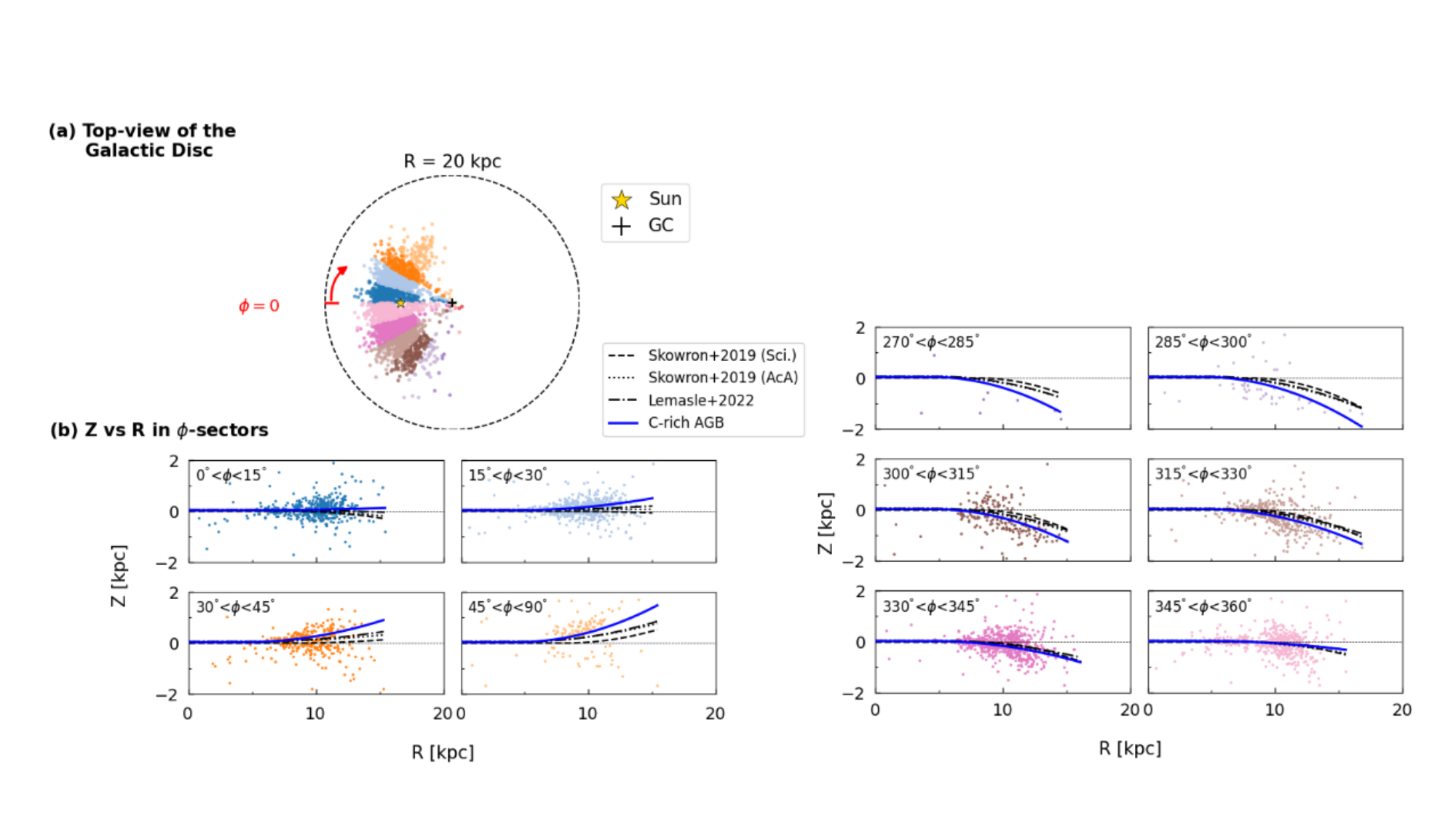} 
\caption{
Demonstration of the Galactic warp traced by C-rich AGB stars as a function of the Galactocentric azimuth ($\phi$). Panel (a) shows a top-down view of the Galactic disc  divided into 10 azimuthal sectors in the Galactocentric polar coordinate system. The azimuth $\phi$=0$^\degree$ points toward the Sun. The Sun is marked with a yellow star, and the Galactic centre is marked with a plus symbol. 
Panel (b) shows the vertical distance $Z$ as a function of the Galactocentric radius $R$ for each $\phi$-sector, using the same colour scheme as in (a). 
The blue solid line shows the best-fitting warp model to the C-rich AGB stars. 
The black dashed, dotted and dash-dotted lines show the warp models derived from Cepheids by \citet{Skowron2019sci}, \citet{Skowron2019acta}, and \citet{Lemasle2022}, respectively. In the azimuthal bins where the upward warp signature is detected ($\phi = 30^\circ$--$90^\circ$), the best fit to the C-rich AGB stars lies above all three Cepheid models. In the bins where the downward warp signature is detected ($\phi = 285^\circ$--$330^\circ$), it lies below. This indicates that larger warp amplitudes are traced by C-rich AGB stars than those traced by Cepheids in both directions.
}
  \label{z_vs_r} 
\end{figure*}

To examine the warp as a function of Galactocentric azimuth ($\phi$), we analyse the C-rich AGB stars in 10 azimuthal sectors, as indicated in Fig.~\ref{z_vs_r}(a). 
For this azimuthal analysis, we focus on C-rich AGB stars because their larger sample size provides better coverage across the range of azimuthal angles.
Fig.~\ref{z_vs_r}(b) shows the $R$–$Z$ distribution of the C-rich AGB stars in each sector. 
The stars in the $\phi$=285$^\circ$--345$^\circ$ range show a downward displacement, reaching $Z\sim$1~kpc at $R=$14~kpc. In the $\phi=15^\circ$--$90^\circ$  range, stars exhibit an upward displacement, reaching $Z\sim$0.7~kpc. In the $\phi=0^\circ$--$15^\circ$ bin, C-rich AGB stars are distributed close to the Galactic mid-plane.

We evaluate the warp amplitude of C-rich AGB stars, using the model developed for Cepheids by \citet{Skowron2019sci}. We adopt their simplified warp prescription:
\begin{equation}
z(R, \phi) = 
\begin{cases} 
-z_0 & R < R_0 \\
-z_0 + z_1 (R - R_d)^2 \sin(\phi - \phi_0) & R \geq R_d 
\end{cases}
\end{equation}
where $z_0$ is a vertical offset, $z_1$ is the primary warp amplitude, $R_d$ is the warp onset radius, and $\phi_0$ is the azimuth of the line of nodes.  \citet{Skowron2019sci} fixed the parameter $R_d$ at 8~kpc; however, in our analysis, we treat $R_d$ as a free parameter.
We fit all four parameters to the C-rich AGB star sample using robust least-squares regression implemented in \texttt{scipy.optimize} \citep{Scipy2020}. Parameter uncertainties are estimated via bootstrap resampling with 1000 iterations.

The best-fitting parameters are: $R_d = 4.4 \pm 0.3$~kpc, $z_0 = 26.3 \pm 10.4$~pc, $\phi_0 = -4 \pm 2^\circ$, and $z_1 = 0.0133 \pm 0.0014$~kpc$^{-1}$. The resulting warp model for C-rich AGB stars is shown in blue in Fig.~\ref{z_vs_r}(b), alongside the three Cepheid-based models from \citet{Skowron2019sci}, \citet{Skowron2019acta}, and \citet{Lemasle2022}.
\citet{Skowron2019acta} introduced an extended model including a second-order warp amplitude term, $z_2$. However, this additional degree of freedom does not converge for our C-rich AGB sample, likely due to the limited number of stars in the inner Galaxy. We therefore adopt the simplified model from \citet{Skowron2019sci}.

%

\section{Discussions}

We used two populations of AGB stars to trace their Galactic distributions: C-rich AGB stars, which represent populations approximately 1~Gyr old, and intermediate-mass O-rich AGB stars, which trace populations roughly 100--300~Myr old. C-rich AGB stars are predominantly found in the outer Galaxy on the Solar System side. In contrast, intermediate-mass O-rich stars are concentrated in the inner Galaxy toward the Galactic Centre, with a more limited spatial extent than the C-rich sample, predominantly spanning approximately $X\sim-10$ to $-4$~kpc.

Previously, \citet{Demers2007} studied the Galactic structure traced by the kinematics of 70 C-rich AGB stars. However, only six stars were located beyond $R > 15$~kpc. \citet{Battinelli2013} expanded this sample by adding rotational velocity measurements for a further 36 C-rich stars and showed that the radial distribution of their rotational velocities is similar to that observed in red-clumps \citep{Momany2006}, indicating the presence of the warp. Using \textit{Gaia} DR3 and a machine-learning classification, we identify 3510 C-rich AGB stars and, for the first time, demonstrate the radial extent of the Galactic warp in this stellar population, as well as measure the vertical ($Z$) displacements as a function of azimuth.

The warp amplitude, or vertical displacement ($Z$), may depend on the stellar population. The typical age of C-rich AGB stars is $\sim$1~Gyr, and we first compare them with younger stars, Cepheids \citep[a few tens to $\sim$300~Myr; ][]{Skowron2019sci}.
In the projected $X$--$Y$ distributions (Fig.~\ref{crich_distributions}), the upward warp of C-rich AGB stars reaches $Z \sim 0.7$~kpc at $R \sim 14$~kpc. This is larger than the Cepheid warp amplitudes of $Z_{\rm up} = 0.53$~kpc \citep{Lemasle2022} and $Z_{\rm up} = 0.65$~kpc \citep{Skowron2019sci} at comparable radii. Similarly, the downward warp reaches $Z \sim -1$~kpc, exceeding the Cepheid values of $Z_{\rm down} = -0.79$~kpc \citep{Lemasle2022} and $Z_{\rm down} = -0.54$~kpc \citep{Skowron2019sci}. Overall, the $Z$ amplitude of C-rich AGB stars appears to be up to a factor of $\sim$2 higher than that of Cepheids.

The azimuth-dependent distribution of C-rich AGB stars exhibits higher amplitudes than the Cepheid models in the ranges $30^\circ < \phi < 90^\circ$ and $285^\circ < \phi < 330^\circ$ (Fig.~\ref{z_vs_r}(b)).
However, direct comparison of primary warp amplitude $z_1$ is not straightforward, as it depends on other model parameters (e.g., $z_0$ and the inclusion of higher-order terms such as $z_2$).

Nevertheless, all Cepheid models shown in Fig.~\ref{z_vs_r}(b) exhibit similar overall shapes and lie below the C-rich AGB warp model. This indicates that the larger warp amplitude observed for C-rich AGB stars is a robust result.

Apart from the amplitudes, the other fitted parameters are broadly in good agreement with those of Cepheid studies. We note that the three Cepheid studies adopt different distance estimates: \citet{Skowron2019sci} and \citet{Skowron2019acta} used period-luminosity relations for Cepheids, while \citet{Lemasle2022} used a combination of period-luminosity relations and mid-infrared photometry. 
As warp amplitude measurements are sensitive to the adopted distance estimates (Appendix~\ref{warp_plx_model}), direct parameter comparison should be made with caution. 
The onset radius $R_d$=4.4$\pm$0.3~kpc is consistent with the Cepheid values of 4.2$\pm$0.1~kpc \citep{Skowron2019acta} and 4.9$\pm$0.3~kpc \citep{Lemasle2022}. The vertical offset $z_0$=26.3$\pm$10.4~pc is closer to \citet{Lemasle2022} ($\sim$26~pc) but smaller than \citet{Skowron2019acta} ($\sim$44~pc). However, the orientation parameter $\phi_0=\textbf{-}4\pm$2$^\circ$ differs from both Cepheid studies ($-21.7$$^\circ$ by \citealt{Skowron2019acta} and $-13.48$$^\circ$ by \citealt{Lemasle2022}). The difference in angle is likely due to the azimuthal coverage of our C-rich AGB sample, which is weighted toward the outer Galaxy in the direction of the Sun.

We also fitted the warp model for C-rich AGB stars using distances computed from direct parallax inversion ($d=1/\varpi$), shown in Fig.~\ref{z_vs_r_plx} (Appendix~\ref{warp_plx_model}), and find that although the warp measurements differ from those obtained with \citet{Bailer-Jones2021} distances, C-rich AGB stars appear to trace larger warp amplitudes than Cepheids regardless of the adopted distance estimate.

Classical Cepheids and C-rich AGB stars trace distinct stellar populations in age and metallicity, and therefore probe different regions of the Galaxy. Classical Cepheids are more massive (3--10$M_{\odot}$; \citealt{Georgy2013}) and younger ($<$400~Myr; \citealt{Skowron2019sci}) than C-rich AGB stars, which have lower initial masses ($\sim$2--2.5~$M_{\odot}$) and ages of $\sim$1~Gyr. The formation of C-rich AGB stars requires efficient third dredge-up episodes, which occur preferentially at sub-solar metallicities \citep{Karakas2007}. Given the negative radial metallicity gradient of the Galaxy \citep{Genovali2014}, C-rich AGB stars are mainly found in the outer Galaxy beyond the Solar circle (Fig.~\ref{crich_distributions}(a)), where metallicities are lower. In contrast, Cepheids are distributed across all Galactic quadrants \citep{Skowron2019sci, Minniti2021}. Cepheids toward the inner Galaxy have been found to be younger, whereas those in the outer Galaxy tend to be older, because metal-poor environments allow lower-mass stars to enter the instability strip \citep{Skowron2019sci, Anders2025}. Consequently, our warp measurement from C-rich AGB stars is well-constrained over certain azimuthal ranges ($\varphi=$0--90$^\circ$ and $\varphi=$285--360$^\circ$), and the observed differences in warp amplitudes may reflect both the stellar properties of the tracer population and the regions of the Galaxy they probe.

Comparison with older stellar populations, such as red clump (RC) stars, is not straightforward. The ages of RC stars span a wide range, from $\sim$1 to 12~Gyr \citep{Wang.2020q4g}, overlapping with the typical ages of C-rich AGB stars ($\sim$1~Gyr).
A warp amplitude of $Z_{\rm up} \approx 1.3$~kpc for RC stars was reported by \citet{Uppal2024} based on analysis in the $Y$--$Z$ projection, which is larger than the value derived from C-rich AGB stars ($\sim$1~kpc). Furthermore, \citet{Khanna2025} suggested the RC amplitude is even larger in the outer region ($|R|>$12 kpc).
However, \citet{Wang.2020q4g} found that the warp is stronger for younger RC populations, suggesting that the relation between age and warp amplitude is not monotonic. 
A complication is that the mass range of RC stars depends on metallicity, with higher metallicity allowing a broader range of initial masses to populate the RC phase \citep{2016ARA&A..54...95G}. This contrasts with C-rich AGB stars, which preferentially form in lower-metallicity environments. Consequently, RC stars and C-rich AGB stars may trace different regions of the Galaxy, and a more detailed and a dedicated, comprehensive comparison is required to disentangle age and metallicity effects from Galactic structure.

The intermediate-mass O-rich stars are less numerous; therefore, comparison with Cepheids is only possible in the Y-Z projection analysis. Although the scatter is large, the downward warp traced by the O-rich AGB stars reaches approximately Z$=-1$ kpc at Y$=-$10 kpc. This appears to be roughly consistent with the warp traced by Cepheids \citep{Skowron2019sci} or younger giants \citep{Poggio.2025}, as expected given their similar ages.

\section{Conclusions}

We demonstrate that AGB stars are highly effective tracers of Galactic structures. With distinct spectral features (TiO and VO bands for O-rich AGB stars, and CN bands for C-rich AGB stars), they can be reliably identified from \textit{Gaia} XP spectra. Their high luminosity allows them, with {\it Gaia} parallaxes, to probe stellar distributions up to $\sim$12~kpc from the Solar system. Combining {\it Gaia} and 2MASS photometry enables the selection of intermediate-mass O-rich AGB stars (main-sequence mass: $\sim$3--5~\Msun, age: 100--300~Myr) and C-rich AGB stars (main-sequence mass: 2--2.5~\Msun, age: 1~Gyr).

AGB stars, particularly C-rich AGB stars, represent an intermediate-age stellar population that complements traditionally used young tracers of the Galactic warp, such as Cepheids and young giants, providing an independent perspective on the warp at intermediate stellar ages.

C-rich AGB stars clearly trace the Galactic warp in the outer Galaxy. The warp shows a clear azimuthal dependence, with a downward displacement of $Z\sim-1$~kpc at $R\sim14$~kpc for $Y < 0$, and an upward displacement of $Z\sim0.7$~kpc at $R\sim14$~kpc for $Y > 0$, consistent with the known morphology of the Galactic warp. For intermediate-mass O-rich AGB stars, a downward warp signature is detected for $Y < 0$, and only a tentative upward signal is seen for $Y > 0$, due to the smaller sample size.

Previous studies have suggested that the warp amplitudes may depend on stellar age. C-rich AGB stars, with typical ages of $\sim$1~Gyr, appear to show larger warp amplitudes than those measured from Cepheids ($<$400~Myr). In contrast, comparisons with older populations, such as red clump (RC) stars, are less straightforward because RC stars predominantly originate from super-solar metallicity populations, whereas C-rich AGB stars are more commonly associated with sub-solar metallicity populations.
These results demonstrate that C-rich AGB stars can serve as effective tracers of the Galactic warp, providing constraints on warp amplitudes at intermediate stellar ages. The observed warp properties reflect both the spatial distribution and intrinsic characteristics  (e.g. age and metallicity) of the tracer population. This highlights the importance of using multiple stellar tracers with complementary coverage to fully characterise the Galactic warp.

\section*{Acknowledgments}

TK acknowledges PhD studentship from the UKRI Centre for Doctoral Training in Artificial Intelligence, Machine Learning \& Advanced Computing (AIMLAC) - Grant Ref: EP/S023992/1. M.M. and R.W. acknowledges support from the STFC Consolidated grant (ST/W000830/1). RW acknowledges support from the Research Ireland Pathway programme under grant number 21/PATH-S/9360. JH acknowledges the support of a UKRI Ernest Rutherford Fellowship ST/Z510245/1. This work has made use of data from the European Space Agency (ESA) mission {\it Gaia} (\url{https://www.cosmos.esa.int/gaia}), processed by the {\it Gaia} Data Processing and Analysis Consortium (DPAC, \url{https://www.cosmos.esa.int/web/gaia/dpac/consortium}). Funding for the DPAC has been provided by national institutions, in particular the institutions participating in the {\it Gaia} Multilateral Agreement. This work has made use of the Python package {\sc GaiaXPy}, developed and maintained by members of the {\it Gaia} Data Processing and Analysis Consortium (DPAC) and in particular, Coordination Unit 5 (CU5), and the Data Processing Centre located at the Institute of Astronomy, Cambridge, UK (DPCI). This publication makes use of data products from the Two Micron All Sky Survey, which is a joint project of the University of Massachusetts and the Infrared Processing and Analysis Center/California Institute of Technology, funded by the National Aeronautics and Space Administration and the National Science Foundation. This publication makes use of data products from the Wide-field Infrared Survey Explorer, which is a joint project of the University of California, Los Angeles, and the Jet Propulsion Laboratory/California Institute of Technology, and {\ it NEOWISE}, which is a project of the Jet Propulsion Laboratory/California Institute of Technology. {\it WISE} and {\it NEOWISE} are funded by the National Aeronautics and Space Administration.

\section*{Data Availability}
 
The sample selected by the machine-learning method will be published alongside the main paper (Kushwahaa et al. submitted).



\bibliographystyle{mnras}
\bibliography{galactic_warp} 




\appendix

\section{Dependence of the Galactic Warp on Distances Estimates}\label{warp_plx}

\begin{figure}
    \centering
    \includegraphics[width=0.95\linewidth]{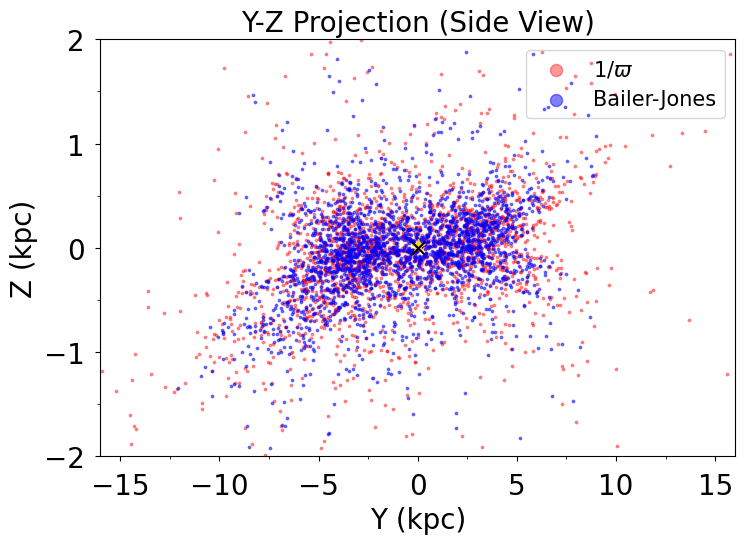}
    \caption{Comparison of distance estimates for the Galactic distribution of C-rich AGB stars, shown from the Galactic centre perspective toward the Solar System. The $Y$ and $Z$ coordinates derived using distances from \citet{Bailer-Jones2021} are shown in blue, while those computed from direct parallax inversion ($d=1/\varpi$) are shown in red. The upward ($Z>0$) and downward ($Z<0$) warp structures inferred from parallax-inverted distances closely resemble those obtained using \citet{Bailer-Jones2021} distances, indicating that the inferred warp morphology is not driven by the adopted distance prior.}
    \label{fig:warp_plx_c}
\end{figure}

\begin{figure}
    \centering
    \includegraphics[width=0.95\linewidth]{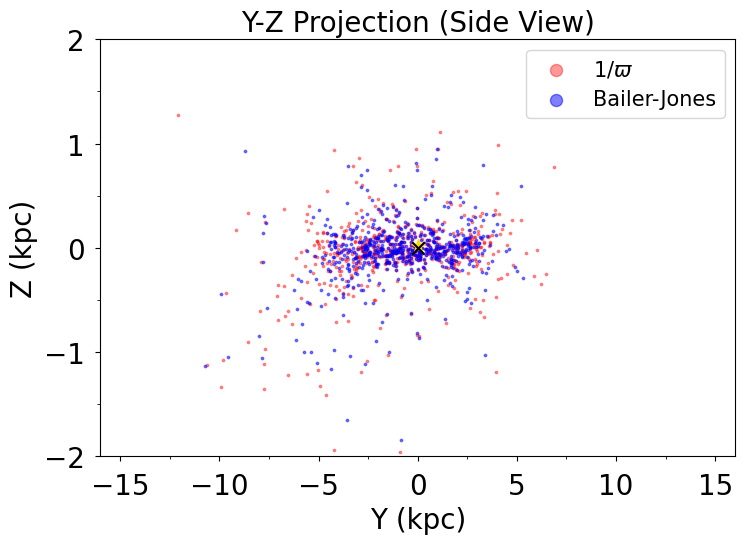}
    \caption{Same as Fig.~\ref{fig:warp_plx_c}, but for intermediate-mass O-rich AGB stars. Both distance estimates show consistent indications of upward and downward warp structures, suggesting that the inferred warp signature is robust against the choice of distances.}
    \label{fig:warp_plx_o}
\end{figure}

The geometric distances from \citet{Bailer-Jones2021} are inferred using a prior of Galactic structure based on the model of \citet{Rybizki2020}, which includes a warp. To assess whether this prior artificially introduced warp, we repeated the analysis using distances derived directly from parallax ($d=1/\varpi$), which are independent of any prior.

We restrict this comparison to stars with high-quality astrometry, adopting RUWE$<$1.4 and $\varpi/\sigma_\varpi>$5. Fig.~\ref{fig:warp_plx_c} shows the resulting $Y-Z$ distribution of C-rich AGB stars for both distances. As \citet{2018A&A...616A...9L} and \citet{2018AJ....156...58B} pointed out, distances derived directly from parallax inversion exhibit larger scatter at greater distances from the Sun. This is because parallax uncertainties arise from multiple sources, are often non-linear, and become amplified when distances are computed by inverting the parallax. Nevertheless, the overall warp morphology, including the upward ($Z>0$) and downward ($Z<0$) displacements found in the parallax-inversion distances, is preserved and remains qualitatively similar to that obtained using \cite{Bailer-Jones2021}'s distances. In particular, the downward warp begins at $Y\sim-6$~kpc and extends to about $Y\sim-14$~kpc, while the upward warp starts at $Y\sim-5$~kpc and extends to $Y\sim-10$~kpc, consistent with  Fig.~\ref{crich_distributions}(d).

Similarly, Fig.~\ref{fig:warp_plx_o} shows the $Y-Z$ projection for intermediate-mass O-rich AGB stars. Both distance estimates reveal a consistent downward warp signature and a tentative upward component.

These tests demonstrate that the detected warp signature in the AGB stars is robust and not an artefact of the adopted distance.

\subsection{Warp model fitting with \texorpdfstring{$1/\varpi$}{1/parallax} distances}\label{warp_plx_model}

\begin{figure}
   \includegraphics[trim={12.5cm 5.3cm 5.5cm 2.2cm},clip, width=0.95\textwidth]{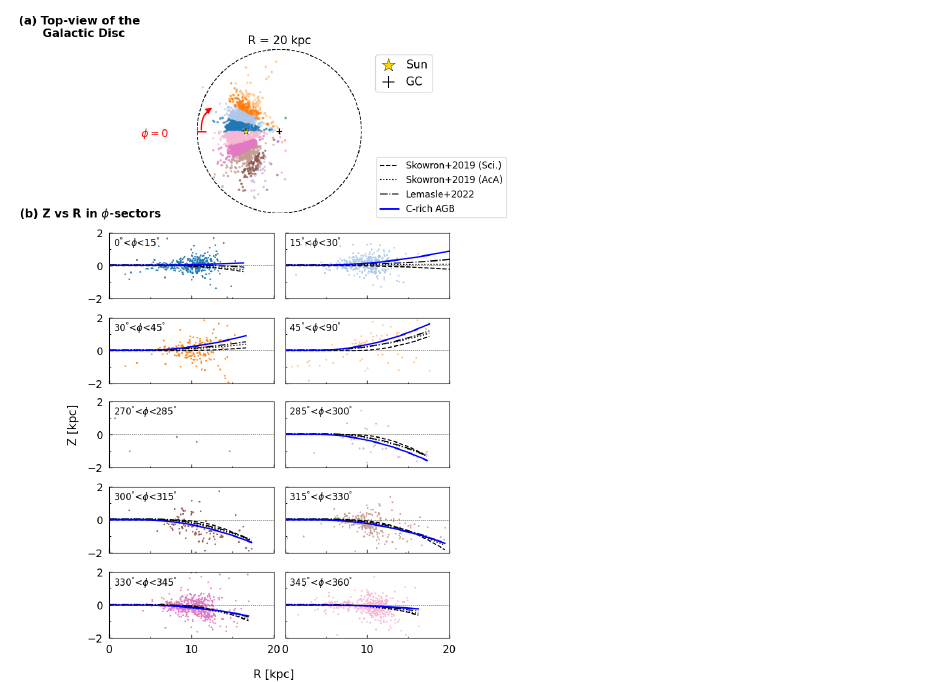} 
\caption{
Same as Fig.~\ref{z_vs_r} but using $1/\varpi$ distances instead of \citet{Bailer-Jones2021} distances. The blue solid line in panel (b) shows the best-fitting warp model to the C-rich AGB stars. The black dashed, dotted and dash-dotted lines show the warp models derived from Cepheids by \citet{Skowron2019sci}, \citet{Skowron2019acta}, and \citet{Lemasle2022}, respectively. 
The C-rich AGB fit exhibits a larger warp amplitude than the Cepheid models, regardless of whether distances are derived from direct parallax inversion or from \citet{Bailer-Jones2021} (Fig.~\ref{z_vs_r}).
}
  \label{z_vs_r_plx} 
\end{figure}

We also tested whether the azimuthal dependence of the warp structure is recovered when using distances derived solely from parallaxes, rather than the distances from \citet{Bailer-Jones2021}.
Fig.~\ref{z_vs_r_plx} shows the $R$--$Z$ distribution of C-rich AGB stars based on $1/\varpi$ distances in the same azimuthal sectors as in Fig.~\ref{z_vs_r}. We fitted all four parameters freely to the distributions of the C-rich AGB stars, following the simplified warp model from \citet{Skowron2019sci}.
The fitted C-rich AGB model traces larger warp amplitudes than all three Cepheid models in the upward bins ($30^\circ<\phi<90^\circ$), and comparable or slightly larger amplitudes in the downward bins ($285^\circ<\phi<330^\circ$) at $R\sim14$~kpc. This trend is broadly consistent with the results obtained using \citet{Bailer-Jones2021} distances (Fig.~\ref{z_vs_r}).

The best-fitting parameters based on the parallax distances are $R_d=3.4\pm0.5$~kpc, $z_0=4.7\pm14.5$~pc, $\phi_0=-2\pm3^\circ$, and $z_1=0.0088\pm0.0014~\mathrm{kpc^{-1}}$. These values differ from those obtained using the distances from \citet{Bailer-Jones2021}. Although the fitted curves appears reasonable in Fig.~\ref{z_vs_r_plx}, the warp amplitude parameter was not well constrained.

\begin{figure}
   \includegraphics[trim={12.5cm 5.3cm 6.cm 2.2cm},clip, width=0.95\textwidth]{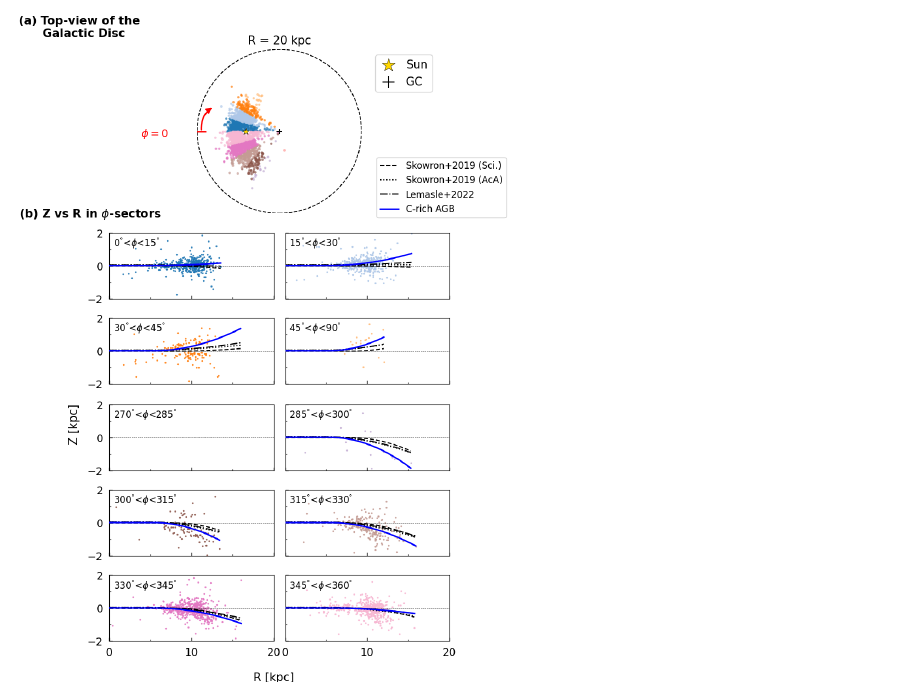} 
\caption{Same as Fig.~\ref{z_vs_r_plx} but using zero-point corrected $1/\varpi_\mathrm{corr}$ distances following \citet{Lindegren2021}. The C-rich AGB fit shows larger warp amplitudes than all three Cepheid models.}
  \label{z_vs_r_plx_corr} 
\end{figure}

We additionally applied the parallax zero-point correction of \citet{Lindegren2021} to account for the systematic bias in \textit{Gaia} DR3 parallaxes. The median zero-point offset for our sample is $-$0.029~mas. We then computed the $1/\varpi_\mathrm{corr}$ distances, and followed the same fitting procedure as Fig.~\ref{z_vs_r_plx}. The best-fitting parameters obtained using zero-point corrected distances are $R_d=5.35\pm0.4$~kpc, $z_0=10.4\pm10.8$~pc, $\phi_0=-1\pm2^\circ$, and $z_1=0.0203\pm0.0035~\mathrm{kpc^{-1}}$. The C-rich AGB warp model in Fig.~\ref{z_vs_r_plx_corr} clearly shows larger warp amplitudes than all three Cepheid models in both the upward and downward warp bins.

All warp models fitted to the C-rich AGB distributions using three distance methods, \citet{Bailer-Jones2021}, $1/\varpi$, and $1/\varpi_\mathrm{corr}$, show that C-rich AGB stars trace larger warp amplitudes than Cepheids, regardless of the adopted distance estimate.

\bsp	
\label{lastpage}
\end{document}